\documentclass[iop,apj]{emulateapj}
\usepackage{amsmath,amssymb,amstext}

\usepackage[breaklinks,colorlinks,citecolor=blue,linkcolor=magenta]{hyperref} 

\usepackage[all]{hypcap} 

\usepackage{aas_macros}
\usepackage{natbib}
\usepackage{color}
\shorttitle{Superburst decomposition}
\shortauthors{Koljonen et al.}

\begin{document}

\title{Evidence of spreading layer emission in thermonuclear superburst}
\author{K. I. I. Koljonen\altaffilmark{1}, J. J. E. Kajava\altaffilmark{2}, E. Kuulkers\altaffilmark{2}}
\affil{$^{1}$New York University Abu Dhabi, PO Box 129188, Abu Dhabi, UAE}
\affil{$^{2}$European Space Astronomy Centre (ESA/ESAC), Science Operations Department, E-28691, Villanueva de la Ca\~{n}ada, Madrid, Spain}

\altaffiltext{1}{email: karri.koljonen@nyu.edu}

\begin{abstract}

When a neutron star accretes matter from a companion star in a low-mass X-ray binary, the accreted gas settles onto the stellar surface through a boundary/spreading layer.
On rare occasions the accumulated gas undergoes a powerful thermonuclear superburst powered by carbon burning deep below the neutron star atmosphere.
In this paper, we apply the non-negative matrix factorization spectral decomposition technique to show that the spectral variations during a superburst from 4U 1636--536 can be explained by two distinct components: 1) the superburst emission characterized by a variable temperature black body radiation component, and 2) a quasi-Planckian component with a constant, $\sim$2.5 keV, temperature varying by a factor of $\sim$15 in flux.
The spectrum of the quasi-Planckian component is identical in shape and characteristics to the frequency-resolved spectra observed in the accretion/persistent spectrum of neutron star low-mass X-ray binaries, and agrees well with the predictions of the spreading layer model by Inogamov \& Sunyaev (1999).
Our result is yet another observational evidence that superbursts -- and possibly also normal X-ray bursts -- induce changes in the disc-star boundary.

\end{abstract}

\maketitle

\section{Introduction}




Low-mass stars in interacting binary configurations with a neutron star (NS) are the most common type among low-mass X-ray binaries \citep[LMXB; see e.g.,][for review]{TvdH06,DGK07}.
The mass lost by the low-mass star is accumulated onto the NS surface through an accretion disc.
In order for the matter to decelerate from Keplerian disk rotation to the slower rotational motion of the NS, a boundary layer must form between the accretion disk and the NS surface. The energy release in this process is substantial and the resulting radiation from the boundary layer can be more luminous than the radiation from the accretion disk \citep{SS86}.
The physical properties, the geometry and the radiation mechanisms of such boundary layers are not well known at present. 
Moreover, depending on the accretion state of the NS-LMXB, the properties and radiation mechanism of the boundary layer can be different. 
In the hard (``island'') spectral state, the optically thick accretion disk may be truncated to larger radii \citep{DGK07} and the optically thin, hot inner flow smoothly connects to the NS through an optically thin boundary layer \citep{DDS01}.
When the accretion rate increases the inner edge of the disk moves inwards and the hot, inner flow collapses/condensates into the disk (e.g. \citealt{MLM07}). 
As the source moves from the hard state to the soft (``banana'') spectral state, the boundary layer becomes optically thick to Comptonization and cools down causing a rapid drop in the X-ray hardness ratio \citep{DGK07}.  
Two distinct geometries of the boundary layer are considered possible: In a classical boundary layer model (hereafter referred to as BL) the rotational velocity decreases over a large radial extent in the disc midplane (e.g., \citealt{Pringle77,SS86,PS01}), or in the alternative spreading layer model (hereafter referred to as SL) the radial extent of the layer is smaller but the gas spreads over a considerable height from the equatorial plane towards higher stellar latitudes (e.g., \citealt{IS99,IS10}). 
Radiative transfer calculations of the BL model by \citet{GS02} predict spectra that are harder than what is observed in NS-LMXBs in the soft spectral state, whereas the SL model calculations by \citet{SP06} predict softer, quasi-Planckian spectra with a color temperature of $2.4$--$2.6$ keV as long as the accretion rate ($\dot{M}$) is sufficiently high ($>$ 10 \% of the Eddington accretion rate).
The SL model predictions are in good agreement with observed values of NS-LMXBs in the soft spectral state (e.g., \citealt{MIK84,LRH07}).    


During accretion episodes NS-LMXBs may show ``superbursts'', which are rare and unusually long thermonuclear X-ray bursts \citep{CHK00}.
Superbursts last several hours whereas the much more commonly observed type I X-ray bursts (see \citealt{LvT93} for review) last only 10-100 seconds. This difference is thought to arise from different burning regimes in the NS atmosphere. 
The type I X-ray bursts are powered by helium and/or hydrogen burning in the NS ``ocean'', while the superbursts are powered by carbon burning at greater depths \citep{CB01}. 
Up to date, only a handful of superbursts have been observed, with only two that have high quality data covering most of the 
burst: 4U 1636--536 \citep{SM02} and 4U 1820--303 \citep{SB02}.
During a thermonuclear burst the X-ray emission in NS-LMXBs can arise from three distinct regions that are all emitting more or less in the same energy band with similar spectral shape: an X-ray bursting NS surface, a BL/SL, and an accretion disc. During an X-ray burst the NS surface can heat up to 3 keV radiating black body-like emission with temperature comparable to that of the BL/SL. If the source is in the soft X-ray spectral state, the accretion disc radiates as a multicolor black body with a temperature of $\sim$ 1 keV, in addition to acting as a reflecting medium to the emission from the bursting NS surface and the BL/SL \citep{Ballantyne2004}. Observationally, the typical NS-LMXB spectrum in the soft state and/or during an X-ray burst is smooth and curved, and can be fitted with several models that include multiple black body-like components with varying temperatures. Usually, the models are degenerate so that the source spectra cannot be fitted unambiguously with a single model, but rather with a variety of models with a comparable fit quality \citep{LRH07}. Thus, this ambiguity can lead to a completely different physical interpretations depending on the choice of the model.                       

In the past, there have been attempts to overcome this ambiguity by using extra information from the timing domain to decompose the constituent components forming the total spectra: first in \citet{MIK84}, and subsequently in \citet{GRM03}, \citet{RG06} and \citet{RSP13}. By using frequency-resolved energy spectra they were able to show that the flux variations on timescales of less than a second are caused by a spectral component of a constant spectral shape varying in normalization. In all of the NS-LMXBs studied, 
the spectral shape of this component was similar and represented the emission from the SL.
Similar ambiguities arise when modeling the X-ray spectra of superbursts.
For example, \citet{KBK14a, KBK14b, KCW2015} used reflection models, in addition to the variable background method \citep{WGP2013, iZGM13}, in fitting the spectra from 4U 1636--536 superburst.
The persistent emission was seen to vary significantly during the superburst, which was interpreted to be caused by an increase of the mass accretion rate due to the Poynting-Robertson drag.
Here, the traditional spectral modeling approach left ambiguities whether the persistent spectrum changed only in flux, or if the persistent spectral shape was also variable during the superburst.

In this paper, we extend the above-mentioned spectral decomposition studies to include a thermonuclear superburst from 4U 1636--536, and instead concentrate on slower time variability (16 s). By using spectral decomposition methods described in \citet{K15}, we show that during the superburst a component identical to the SL seen in the persistent emission can be found contributing a sizable fraction of the total luminosity. Smoothly evolving emission from the SL naturally explains the light curves and spectral evolution of the superburst. 


\section{Observations and methods}

We used data from the Proportional Counter Array (PCA; \citealt{JMR06}) onboard Rossi X-ray Timing Explorer taken during the superburst from 4U 1636--536 \citep{SM02} on 22 February 2002. 
We extracted the Standard 2 PCA spectra from observation IDs 50030-02-08-01 and 50030-02-08-02, in intervals of 16 seconds using \textsc{heasoft} 6.17. 
We used PCUs 0, 2 and 3  
and extracted spectra from all three detector layers.
The exposure times were corrected for deadtime effects, and 
we used one background and response per orbit produced by the tools \textsc{pcabackest} and \textsc{pcarsp}, respectively.
The spectra were restricted to span 3--18 keV range to ensure positive values in all channels throughout the superburst for the decomposition procedure, but for the spectral fitting we used a slightly wider band of 3--20 keV. The fitting was conducted using \textsc{isis} \citep{HD00}, and 0.5\% systematic error was introduced to the PCA data. The errors of the resulting fit parameters are quoted at 90\% confidence level. 





\subsection{Spectral decomposition}
Unsupervised spectral decomposition methods have been proven to be a powerful tool in separating a set of X-ray spectra from X-ray binaries and active galactic nuclei into subcomponents \citep{VF04,MPJ06,KMH13,PFM15,K15,DKC16}. In general, the X-ray spectra is decomposed to its constituent components by using matrix factorization techniques, e.g. principal component analysis. In these techniques a source matrix, $X_{ji}$, consisting of a discrete set of spectra with flux values for each energy $j$ and each spectrum $i$, can be linearly decomposed to a mixture of separate source signals $S_{ki}$ weighted over the energy bands by a weight matrix $W_{jk}$ in such a way that $X_{ij} \approx \sum_{k} W_{jk}S_{ki}$, where $k$ is a dummy variable denoting the running number of components in the decomposition. Usually, the number $k$ is small, and only a few components are needed to explain the variability in the data set within data errors. In practice, we can think the weights of a single component as its spectral shape, which has a variable amplitude denoted by the corresponding signal. 

In this paper, we use non-negative matrix factorization (NMF; \citealt{PT94,LS99}) to decompose the data from the thermonuclear superburst of 4U 1636--536. NMF was found to perform best from a collection of linear decomposition methods in disentangling different spectral components in simulated X-ray spectra \citep{K15}. As spectral evolution typically contains non-linear effects in LMXBs, such as the changing temperature of the black body radiation, this will cause the NMF to estimate this effect as a collection of multiple linear components. However, by summing these components together, we are able to estimate the non-linearity present in the spectra (see simulation studies in \citealt{K15}). In the following, the NMF technique is briefly described, and the reader is referred to \citet{K15} for more detailed discussion. 

In NMF, the matrices $W$ and $S$ are found by minimizing a cost function (generalized Kullback-Leibler divergence) under the constraint that they must be non-negative. For calculating the decomposition we used the package \textsc{nmf} \citep{GS10} that calculates the standard NMF \citep{BTG04} by picking random starting values for $W$ and $S$ from a uniform distribution [0,max($X$)] and then updating iteratively 10000 times to find a local minimum of the cost function with a multiplicative rule from \citet{LS01}. The minimization process is repeated for 300 starting points to ensure that the algorithm does not get stuck in a local minimum. Despite this, we found that the solutions have a certain amount of scatter in the resulting NMF components and thus we ran the analysis several times to probe the effect of this scatter (see below).  

To determine the degree of factorization, $k$, we use the $\chi^{2}$-diagram method devised in \citet{K15}. In general, we expect the quality of the factorization, i.e. its similarity with the original data, to be an increasing function of $k$. We aim for a value of $k$, which provides substantially better approximation than nearby smaller values, but only a slightly worse approximation than nearby larger values. In the $\chi^{2}$-diagram method, the reduced $\chi^{2}$-values are calculated between each individual spectrum in the data set with associated errors and the factorization $\sum_{k} W_{jk}S_{ki}$. Then a median is taken from all the $\chi^{2}$-values for a particular $k$:
\begin{equation}
  \chi^{2}_\textrm{red}(k) = \mathrm{Mdn} \Bigg\{ \frac{\sum_{i}[(X_{ji}- \sum_{k}W_{jk}S_{ki})/\sigma_{ji}]^{2}}{\mathrm{max}(j)-k} \Bigg\} .
\end{equation}
This produces a quality measure of how well a particular factorization with a degree $k$ fits in to the data and takes into account the number of components up to those that vary above the noise level. 

\section{Results}

\subsection{NMF components}

\begin{figure}
  \epsscale{1.00}
  \plotone{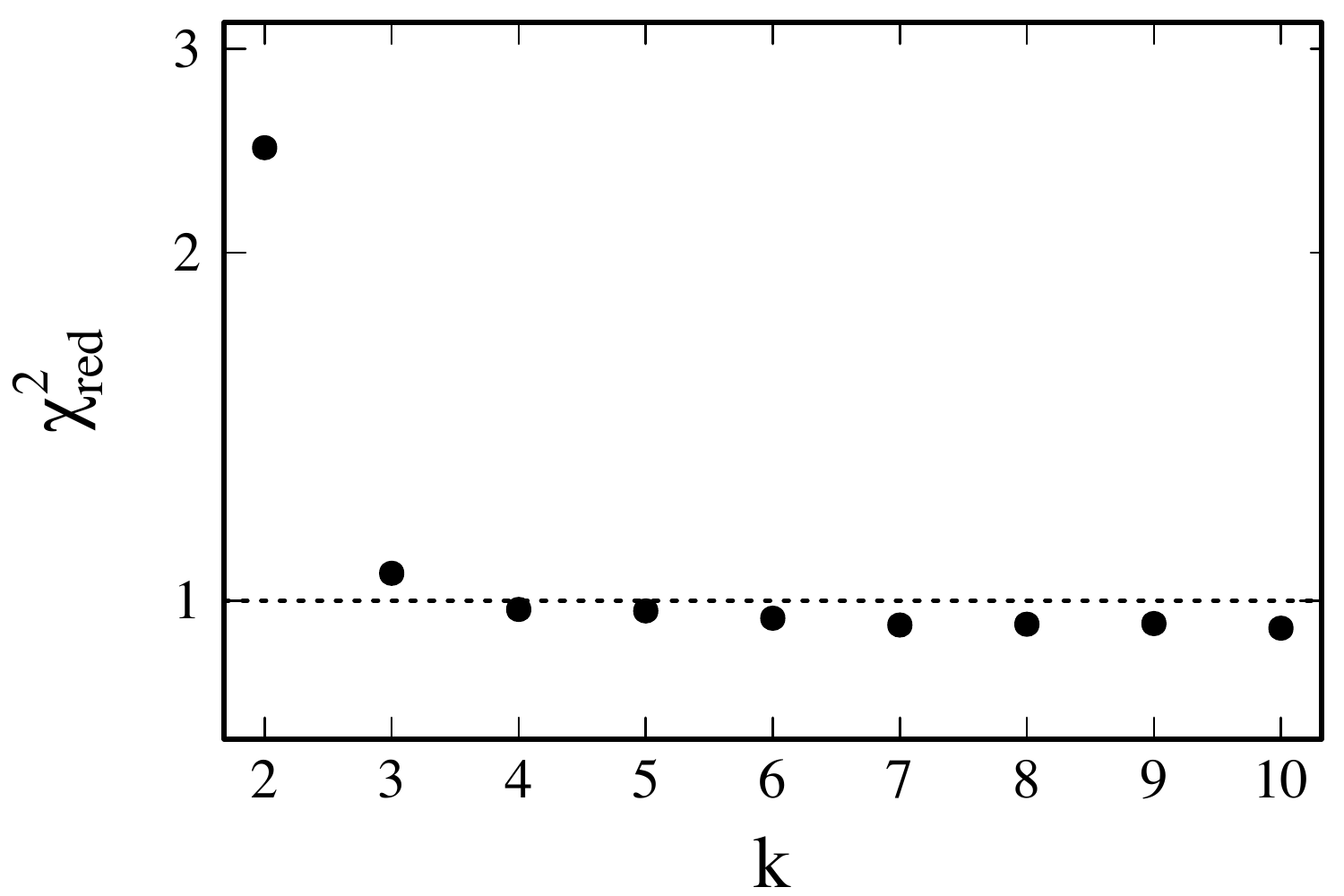}
  \caption{Determining the degree of factorization for NMF analysis. The figure shows the $\chi^{2}$-diagram of the superburst data set from 4U 1636--536. After k=3, the $\chi^{2}$-diagram achieves a value where further increasing the degree of factorization reduces the $\chi^{2}_\textrm{red}$-value only by a small amount (i.e. a kink or an elbow in the diagram). Three components are enough to explain the spectral variability of 4U 1636--536 during the superburst.}
  \label{fac}
\end{figure}

We begin by calculating the $\chi^{2}$-diagram for the superburst data set of 4U 1636--536 (Fig. \ref{fac}). As mentioned above, the chosen degree of factorization should be a point where the quality measure changes from steep to shallow (i.e. a kink or an elbow in the diagram). In addition, this value should be close to 1, so that the decomposition would portray faithfully the original spectra. We identify that three components are enough to explain the variability in the data during the superburst. 

\begin{figure}
  \epsscale{1.17}
  \plotone{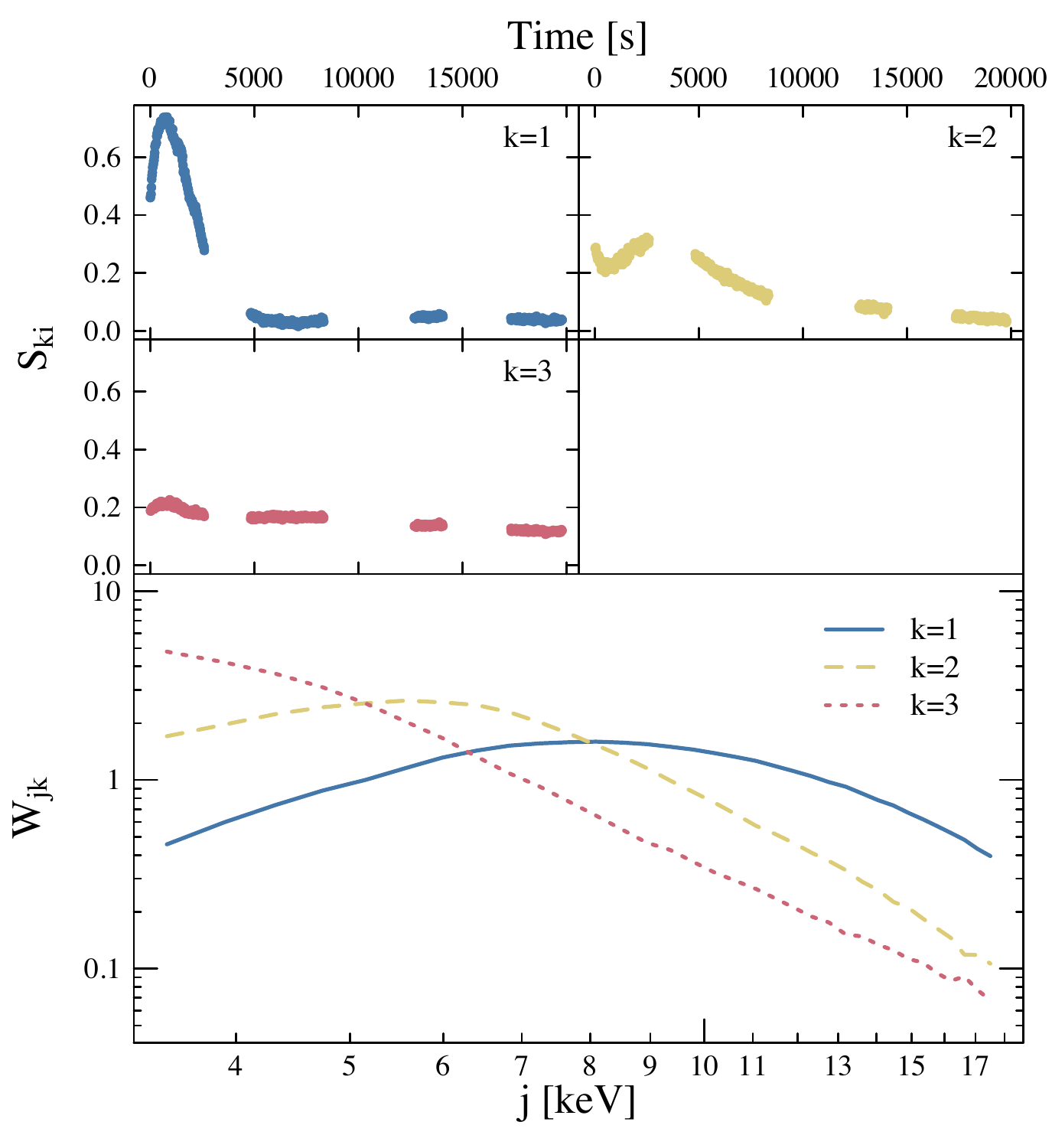}
  \caption{The three NMF components that explain most of the spectral variability in the 4U 1636--536 superburst data set. The smaller top panels show the source signals $S_{ki}$, and the larger bottom panel the weights $W_{jk}$ of the decomposition. The $k=1$ component likely represents the SL.} 
  \label{nmf}
\end{figure} 

\begin{figure}
  \epsscale{1.17}
  \plotone{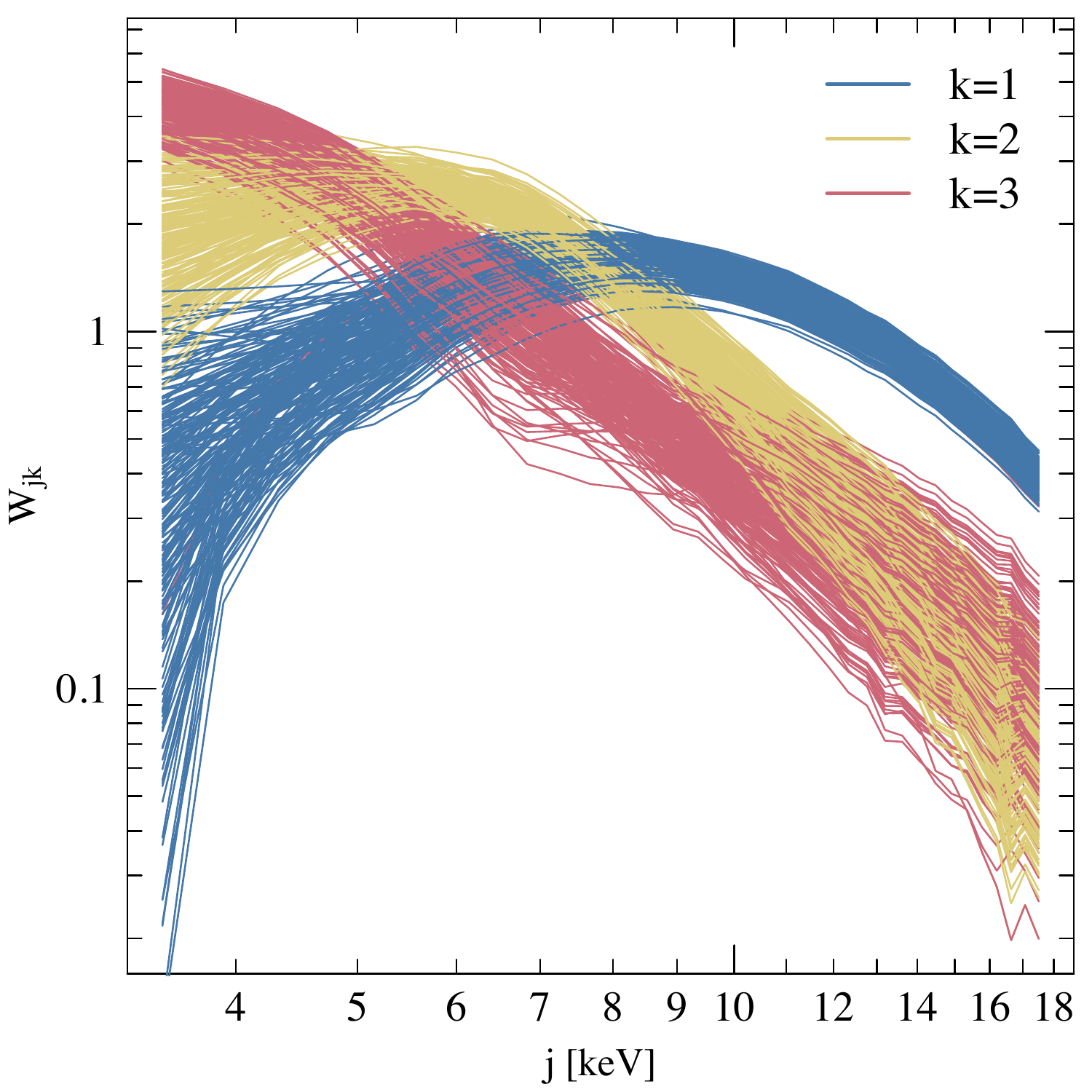}
  \caption{A sample of weights from multiple NMF runs on the superburst data set of 4U 1636--536. The weights vary slightly from run to run, but the overall shape is retained.} 
  \label{mc}
\end{figure} 

Fig. \ref{nmf} shows the signals and weights of one NMF run for three components. The smaller upper panels show the source signals $S_{ki}$ and the bottom panel the weights $W_{jk}$. Note that the ordering of the NMF components is random and does not portray any information. 
The $k=1$ most likely represents the emission component from the SL, and the other components are likely caused by the cooling burst spectra. As the NMF relies on linear decomposition, it is suited best for finding the variability of a spectral component that vary in normalization. As mentioned above, the non-linear effects can be estimated by summing two or more components together that form the spectral component presenting non-linear behavior; in this case the cooling burst spectrum.     

We ran the NMF analysis 200 times to look for deviations in the solutions of the decomposition. Fig. \ref{mc} shows the weights of the resulting NMF components. We note that some amount of scatter is present in the weights of individual NMF runs. However, the weights across the different energy bands, i.e the ``shape of the spectrum'', are fairly consistent in all of the runs. Thus, there exists a small amount of ambiguity in the individual NMF solutions which we take into account in the analysis below.        


\subsection{Spectral components}

\begin{figure*}
  \epsscale{1.17}
  \plotone{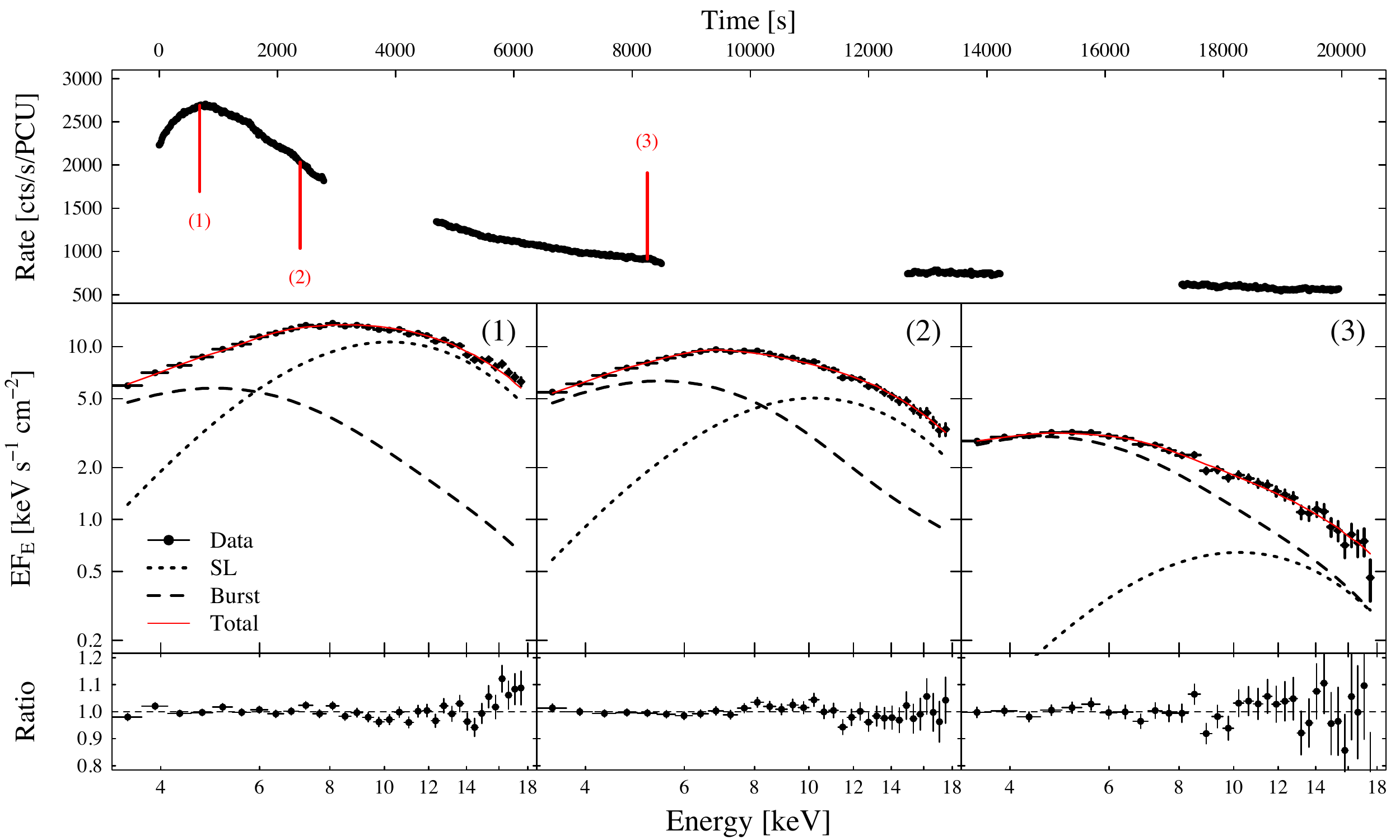}
  \caption{RXTE/PCA lightcurve from the 22nd Feb 2002 superburst of 4U 1636-536 (top panel). Selected spectra from three parts of the superburst decomposed to their corresponding NMF components (middle panels). The ratio of the selected spectra and the NMF decomposition (bottom panels).}
  \label{comp1636}
\end{figure*} 

To interpret the NMF components, we start by reconstructing spectral components from the NMF components we propose as representing the SL ($k=1$), and the cooling burst ($k=2,3$). This can be simply done by combining the calculated weight matrix and signal to form the constructed SL spectra as $M_{SL} = W_{jk}S_{ki}$, where $k=1$, and for the burst spectra as $M_{B} = \sum_{k} W_{jk}S_{ki}$, where $k=2,3$. In Fig. \ref{comp1636}, we show three spectra along the superburst ($\sim$ 700 s, 2400 s, and 8200 s after the start of the burst; top panel), together with their decomposition  to $M_{SL}$ and $M_{B}$ corresponding to the NMF run shown in Fig. \ref{nmf}. During the peak of the superburst (left, middle panel), the SL component dominates the total spectrum, but around 2000 s after the start of the burst (center panel) both components contribute roughly the same amount to the total spectrum, and from there on the burst component starts to dominate the total spectrum (right, middle panel).  

For each NMF run, we construct the $M_{SL}$ and $M_{B}$, and calculate the average spectra, $\bar{M}_{SL}$ and $\bar{M}_{B}$, out of all runs. The errors for the averaged spectra are estimated as a square root of the unbiased sample variance: $\sigma^{2}=1/(n-1)\sum_{i=1}^{n}(M_{i}-\bar{M})^{2}$. These spectra are then imported to \textsc{isis} for spectral fitting. 

\begin{figure}
  \epsscale{1.17}
  \plotone{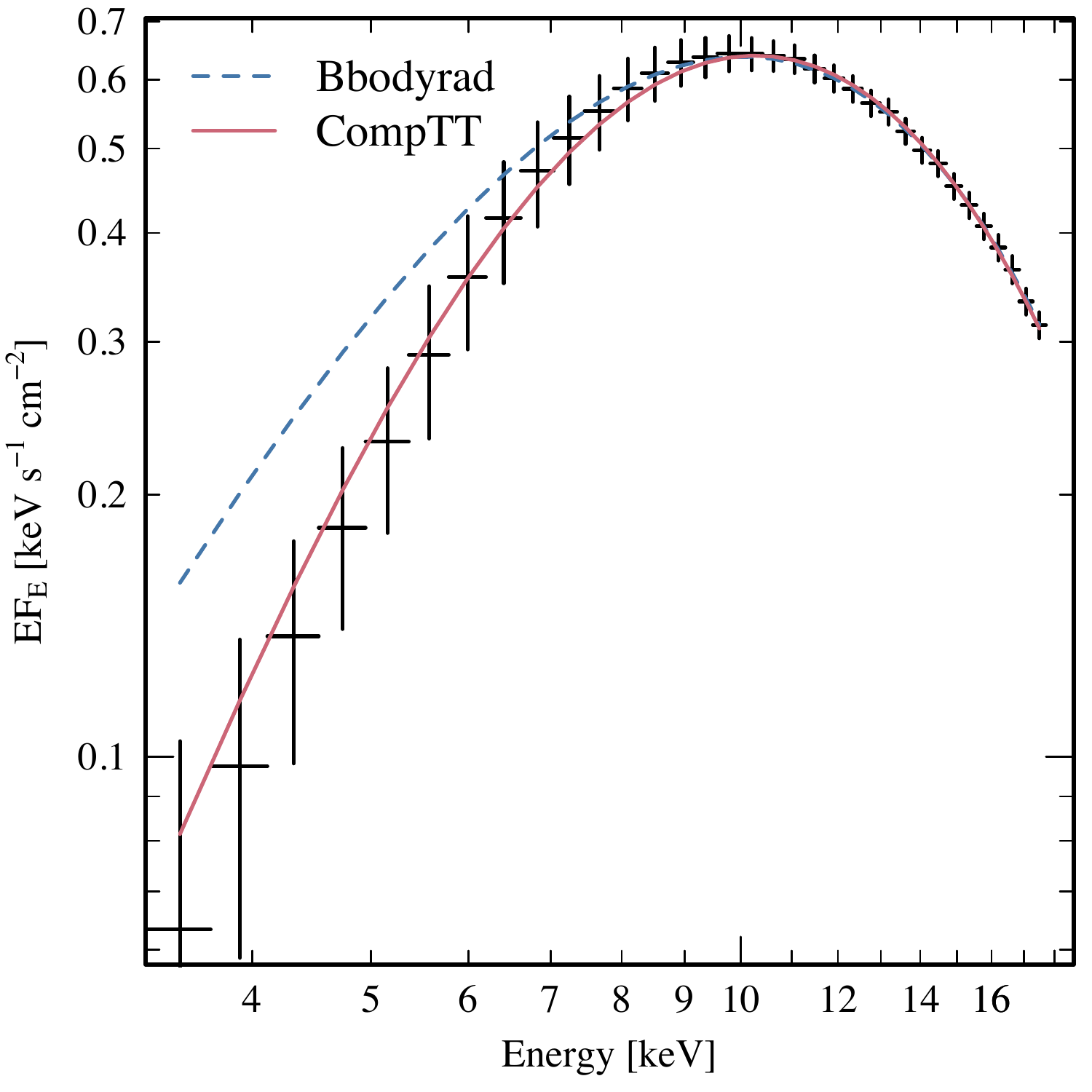}
  \caption{The SL spectrum from 4U 1636--536 fitted with the best-fit saturated Comptonization model (solid red line) and black body radiation model as fitted to the data over 6 keV (dashed blue line). Both models converge over 10 keV but deviate in lower energies.}
  \label{sl}
\end{figure}

For the averaged SL spectrum, we found that above 6 keV it is well fitted with a black body model (\textsc{bbodyrad}) of a temperature 2.56 keV, but in lower energies the SL spectrum is underluminous similar to the more detailed calculation of the spectrum from SL by \citet{SP06}. A better fit is obtained when using a saturated Comptonization model (\textsc{comptt}; \citealt{T94}) with the temperature of the seed photons $kT_{s}=1.5\pm0.2$ keV, and the temperature of the Comptonizing electrons $kT_{e}=2.56\pm0.05$ keV, while fixing the optical depth to $\tau=20$. The fit is not very sensitive to a particular value of $\tau$ as long as it is fairly high, i.e. optically thick. Fig. \ref{sl} shows the difference between the black body and the Comptonization model as fitted to the averaged SL spectrum taken from the peak of the superburst. Similarly, in \citet{RSP13} the frequency-resolved energy spectrum from the SL component was found to be well-fitted with a saturated Comptonization model.

For the averaged burst spectrum, we found that it can be fitted with a cooling black body model (\textsc{bbodyrad}) with an additional spectral component in the higher energies. This component can be modeled with a power law, but also the above SL model result in equally good fits. If the SL spectral component varies in concert with the burst spectra, it is possible that it has ``leaked'' partly to the burst spectra.   

\subsection{Spectral fitting}

\begin{figure}
  \epsscale{1.2}
  \plotone{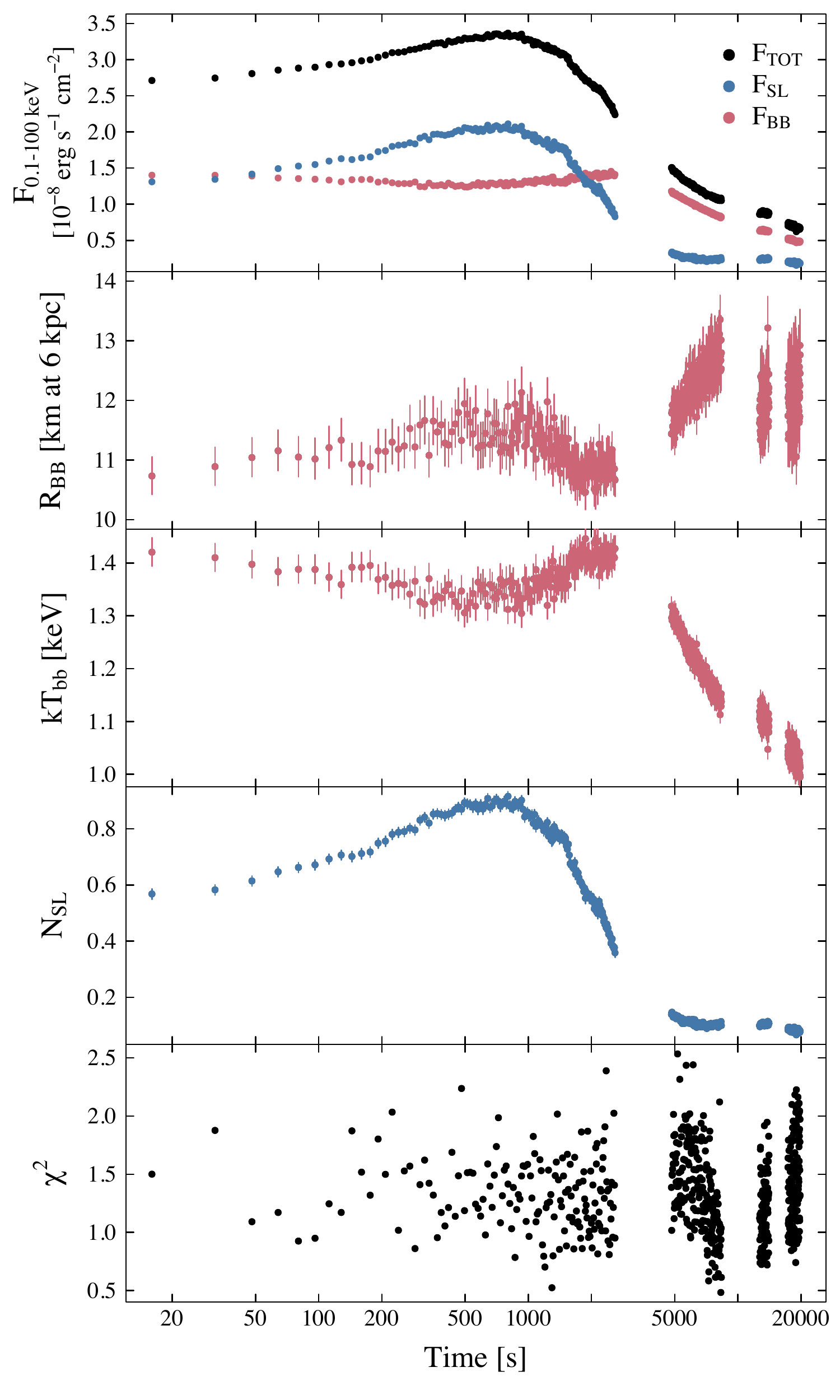}
  \caption{Component fluxes, spectral parameters, and the quality of the spectral fits of the 16 s superburst spectra. From top to bottom: unabsorbed, bolometric (0.1-100 keV) fluxes of the black body (BB), Comptonization (SL) and the total (TOT) spectra, the effective radius in km (assuming a distance of 6 kpc) derived from the black body normalization, the temperature of the black body component, the normalization of the Comptonization component (other parameters are fixed, see text), and the reduced $\chi^{2}$ of the spectral fits, respectively.}
  \label{params_1636}
\end{figure} 

In order to self-consistently verify the NMF decomposition and the above spectral model, we fit the original data with the following model: \textsc{phabs $\times$ (comptt + bbodyrad)}. The parameters of the \textsc{comptt} component are fixed to the values mentioned above, with the normalization left free. The absorption column density is fixed to 0.38 $\times 10^{22}$ cm$^{-2}$ \citep{PKC08}. Thus, this leaves three free parameters in the model: the normalizations of the black body and Comptonization components, and the temperature of the black body radiation. I.e. these can be thought as three degrees of freedom mirroring the NMF result that three components are enough to explain most of the spectral variability. 
The evolution of these parameters and corresponding fluxes throughout the superburst are displayed in Fig. \ref{params_1636}. The top panel shows the unabsorbed, bolometric (0.1-100 keV) fluxes from the black body (BL) and Comptonization (SL) components together with the total bolometric flux (TOT). The middle panels show the apparent black body radius at 6 kpc \citep{GPM06} calculated from the black body normalization, the temperature of the black body, and the normalization of the \textsc{comptt} model, respectively. The bottom panel shows the reduced $\chi^2$ values of the spectral fits. The flux from the SL component decreases by a factor of $\sim$15 during the superburst and at the brightest contributes more than 60\% of the total flux. The black body flux (mirroring the evolution of the black body temperature) is fairly constant at the peak of the burst exhibiting a little dip, and then decreases by a factor of 3 as the burst cools from $\sim$1.4 keV to $\sim$1.0 keV. During the superburst the effective NS radius stays at a fairly constant value of $\sim$ 12 km, but during the burst decay a clear trend can be seen where the radius increases from 11 km to 13 km.

\begin{figure}
  \epsscale{1.2}
  \plotone{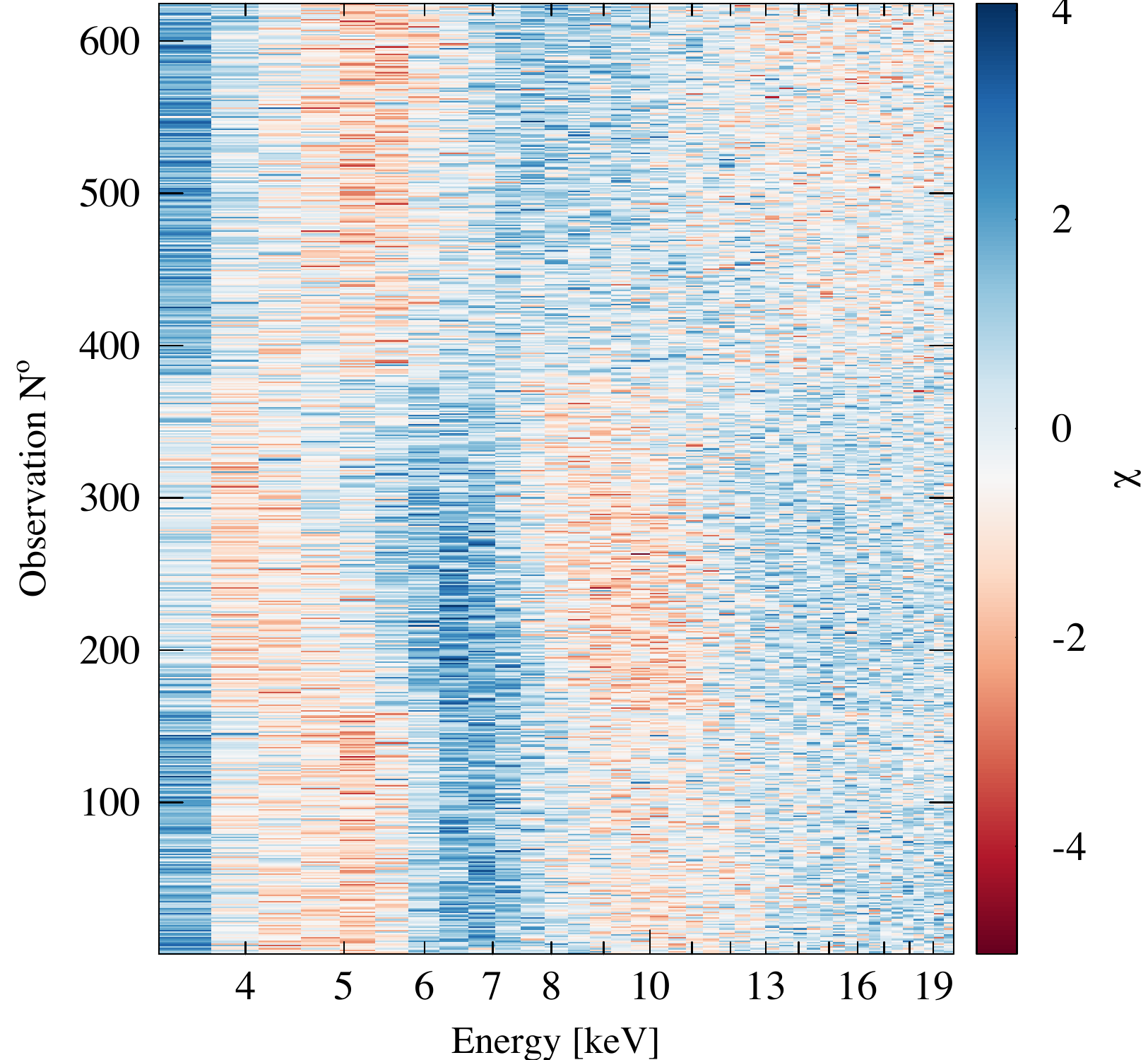}
  \caption{Residuals of the spectral fits shown in Fig. \ref{params_1636}. There are indications of iron features in the form of an iron line at $\sim$6.4 keV, and a blend of iron edges around $\sim$9 keV.}
  \label{residuals_1636}
\end{figure} 

Previous studies have shown \citep{KBK14a,KBK14b} that the burst spectrum is accompanied by iron features in the form of an iron line around 6.5 keV, and an iron edge (or a blend of iron edges) around 9 keV. 
We found that the 16 s resolution data does not prominently show line or edge features, and we can adequately fit all spectra with the above model. However, the fit residuals (Fig. \ref{residuals_1636}) show features around 4--10 keV energy range that are likely caused by the iron features. In this paper, we are more interested in the continuum components, and thus we leave the detailed spectral modeling to future publication (Kajava et al., in prep.).  


\begin{figure}
  \epsscale{1.2}
  \plotone{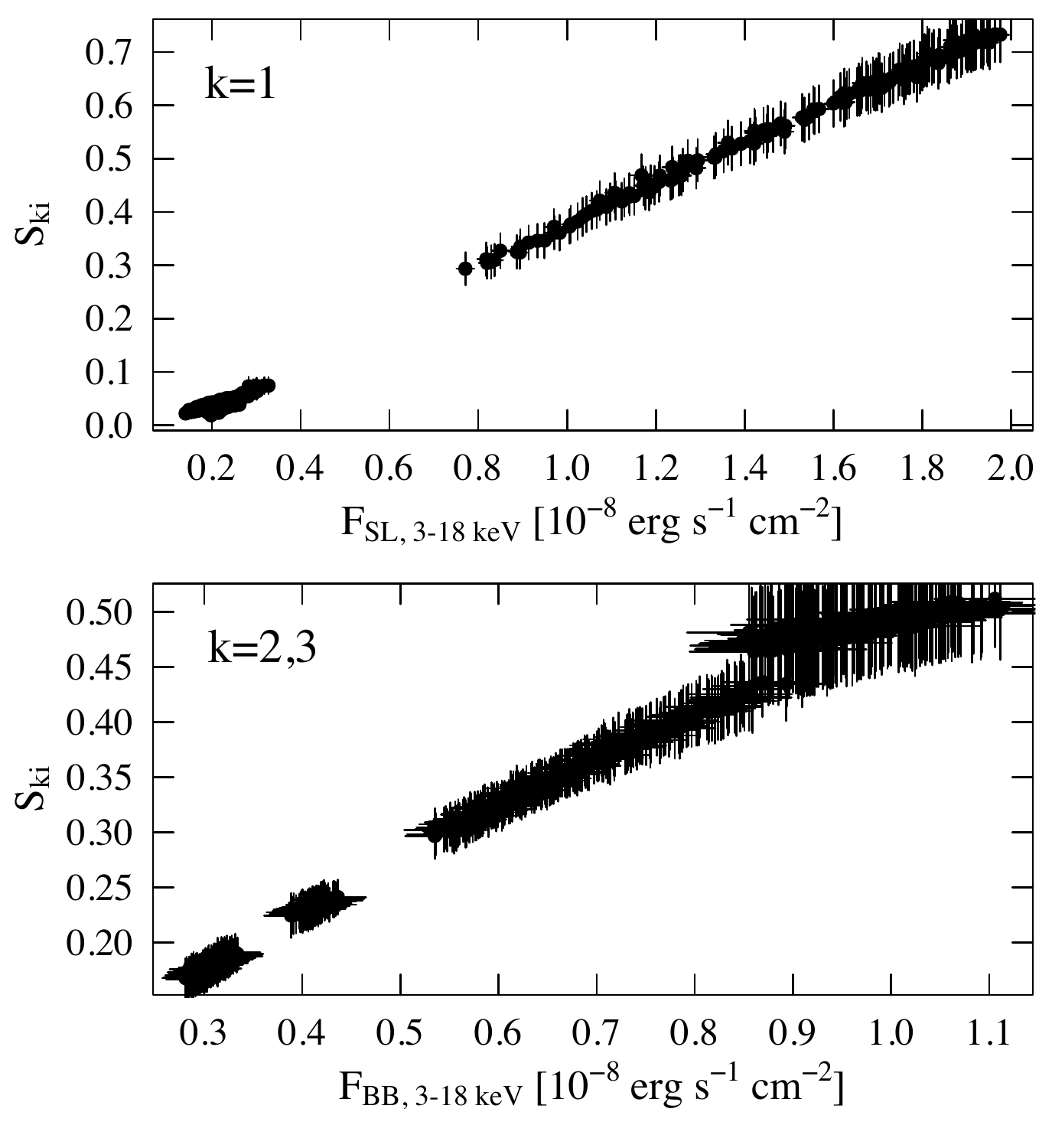}
  \caption{Comparing the spectral component fluxes of the SL and black body to their corresponding, averaged NMF signals (S$_{k=1}$, and S$_{k=2,3}$) from all NMF runs. Note that there are deviations from one-to-one correspondence probably due to our use of simple spectral model, component leaking in the NMF decomposition, or the inadequacy of the three component NMF to estimate some (small) spectral features.}
  \label{compare}
\end{figure}

Finally, we compare the black body and SL component fluxes to their corresponding NMF signals. Fig. \ref{compare} shows the averaged signals from the NMF runs plotted against the component fluxes from the spectral analysis. A linear relation is evident with slight deviations either from our use of a simple spectral model, component leaking in the NMF as discussed above, or that some (small) spectral features are not represented by the three NMF components (the $\chi^{2}_{red}$ is lower at $k=4$, see Fig. \ref{fac}). Thus, we can state that the spectral modeling agrees with the NMF analysis. 



\section{Discussion}



Our results indicate that during an X-ray superburst there are at least two variable components: the cooling X-ray burst emission and a quasi-Planckian component with a constant temperature of 2.4--2.6 keV. The fact that the NMF component $k=1$ (SL) does not break up into several components, and that the spectra can be fitted by imposing a constant temperature ($\sim2.5$ keV) saturated Comptonization model, just as the SL atmosphere calculations by \citet{IS99,SP06,IS10} predict, and what is seen in the persistent (accretion) emission in atolls and Z-sources \citep{GRM03,RG06,RSP13}, suggest that a variable SL is present also during the superburst and has a major contribution to the total X-ray spectra. Therefore, our finding is yet another supporting observational evidence for the SL model in NS-LMXBs. 

Previously, it has been noted that the X-ray burst spectra are statistically much better described if -- in addition to a black body model describing the burst emission -- another variable component is added to the model \citep{WGP2013, iZGM13}. The first and simplest approach is to model the spectrum prior to the onset of the X-ray burst, and allow its flux to vary during the X-ray burst by multiplying the model with a constant (labeled as $f_\textrm{a}$ term). This variable $f_\textrm{a}$-method was applied to the same superburst data used in this paper by \citet{KBK14a, KBK14b, KCW2015}. We note, however, that none of the individual NMF components could be fitted with the model (\textsc{cutoffpl}) that was used to describe the persistent spectrum prior to the X-ray burst in \citet{KBK14a, KBK14b, KCW2015}. Rather, the persistent spectrum can be equally well described by a {\sc diskbb} $+$ {\sc comptt} model, where the {\sc comptt} model component takes the same values as during the superburst. We find that the component changing in normalization (i.e. flux) is the one corresponding to the SL. Thus, the main difference to the previous spectral modeling in \citet{KBK14a, KBK14b, KCW2015} is that the SL spectrum has more flux in higher energies (above 10 keV) than the cutoff powerlaw spectrum as fitted to the persistent spectra. 
For this reason, in our decomposition the burst spectrum appears much cooler during the superburst peak ($\sim$1.4 keV as compared to $\sim$2.5 keV), as the spectrum is now dominated by the SL spectrum at higher energies, but there is need for more flux in the soft X-rays. 
Furthermore, our spectral fits result in comparable black body radii ranging from $\sim11$ km in the superburst peak to $\sim12$ km during the latter three spacecraft orbits, whereas the $R_\textrm{bb}$ values during the peak in \citet{KBK14a} were much lower ($\sim5$ km). 
These changes may be caused by the SL getting slightly wider, effectively blocking the direct burst emission in the first orbit.

The increase of the persistent emission during X-ray bursts has been interpreted as a momentary increase of the mass accretion rate due to the Poynting-Robertson drag \citep{WM89,W92,ML96}. Similarly, during the superburst, one could attribute the increase of the SL emission flux to a 15-fold increase in $\dot{M}$. However, this interpretation suffers from one important drawback. If $\dot{M}$ were to increase by such a large amount, the accretion disc flux should also have increased in concert by the same amount. Assuming that half of the persistent luminosity originates in the accretion disk, the disk flux should contribute a sizable fraction of the total flux during the superburst, in addition to the disk becoming hotter. This is not observed; there is no NMF component (or a sum of components) that can be described by a disk black body model ({\sc diskbb}) with a variable temperature and a constant normalization that would be expected in this case. An alternative scenario is that the changes in the SL flux are instead caused more directly by the superburst. As speculated in \citet{SPR11} and \citet{KNL14}, the X-ray burst occurring underneath the SL can provide additional radiation pressure support to the SL that is ``levitating'' above the stellar photosphere \citep{SP06}. As the SL is radiation pressure supported, the burst emission may push the SL towards higher latitudes during the burst, increasing its luminosity as more burst photons are reprocessed in it, thus mimicking an increase in the accretion rate. 

In a recent work by \citet{DKC16} a similar NMF analysis as done in this paper showed that the persistent spectrum changes during an X-ray burst in the hard state as well. As the burst occurred in the hard state, the SL is likely optically thin and joins smoothly to the optically thin accretion flow, which can be described by a power law spectral model. The slope of the powerlaw component changed during the burst, which is likely caused by a decrease of the equilibrium electron temperature in the hot flow when the external photons from the X-ray burst enter the corona \citep{JZC15}. This is similar to the black hole X-ray binaries in the case where the variable amount of external disc photons may enter the hot flow in the hard state \citep{PV09, MB09}. It seems therefore that the NMF technique has now revealed two possible physical origins to persistent spectral changes during X-ray bursts: variable electron temperatures during the hard state bursts and burst induced increase of the SL emission in the soft state bursts.

It remains to be studied whether the superburst from 4U 1820--303 -- and also normal type-I X-ray bursts -- behave the same way as the 4U 1636--536 superburst. 
However, type-I X-ray bursts cool on much shorter time scales, which reduces the number of good quality spectra per burst, thus making the NMF decomposition not as robust as is the case with a superburst.

\section{Conclusions}

Spectral degeneracy, i.e. the ability to fit a variety of models to same spectrum with comparable quality, is a real problem in X-ray astronomy, which arises either from photon-starved spectra in faint sources, and/or smooth and curved spectra in bright sources. Especially good example can be found in NS-LMXBs, where the spectrum during an X-ray burst and/or soft state consists of multiple black body(-like) components with varying temperatures. This ambiguity has led to different interpretations of the constituent spectral components depending on the fit results. We have shown that by using non-negative matrix factorization, the superburst spectra of 4U 1636-536 can be decomposed into two variable spectral components: the cooling burst spectrum, and a boundary/spreading layer, which is found to be contributing a sizable fraction of the total luminosity during the superburst. The spectral properties of the boundary/spreading layer component favors the spreading layer model \citep{IS99,IS10,SP06}, where the spectrum is a constant $\sim$2.5 keV, quasi-Planckian component varying just in normalization as the burst evolves. This component is also very reminiscent of the frequency-resolved spectral component of a constant spectral shape that is responsible for the sub-second variability in many NS-LMXBs. Smoothly evolving emission from the spreading layer naturally explains the light curves and spectral evolution of the superburst, without the need to invoke a sudden increase of the mass accretion rate (with the Poynting-Robertson mechanism).

\acknowledgments{
We thank Valery Suleimanov, Juri Poutanen and Guobao Zhang for useful discussions.
KIIK and JJEK acknowledge support from the Faculty of the European Space Astronomy Centre (ESAC). 
JJEK acknowledges support from the ESA research fellowship programme.
}

\bibliography{superburst}

\end{document}